\documentclass[aps,10pt,floatfix,reprint,superscriptaddress,nofootinbib]{revtex4-2}

\usepackage{amssymb, amsmath}
\DeclareMathOperator*{\argmin}{arg\,min}
\usepackage[english]{babel}
\usepackage[utf8]{inputenc}
\usepackage{blindtext}
\usepackage{graphicx}
\usepackage{hyperref}
\hypersetup{
    colorlinks,
    linkcolor={red!50!black},
    citecolor={blue!50!black},
    urlcolor={blue!80!black}
}
\usepackage{physics}
\usepackage{xcolor}
\usepackage[normalem]{ulem}
\usepackage{float}

\renewcommand{\vec}[1]{\mathbf{#1}}

\begin{document}

\title{Message-Passing Neural Quantum States for the Homogeneous Electron Gas}
\author{Gabriel Pescia}
\affiliation{Institute of Physics, École Polytechnique Fédérale de Lausanne (EPFL), CH-1015 Lausanne, Switzerland}
\affiliation{%
Center for Quantum Science and Engineering, \'{E}cole Polytechnique F\'{e}d\'{e}rale de Lausanne (EPFL), CH-1015 Lausanne, Switzerland
}

\author{Jannes Nys}
\affiliation{Institute of Physics, École Polytechnique Fédérale de Lausanne (EPFL), CH-1015 Lausanne, Switzerland}
\affiliation{%
Center for Quantum Science and Engineering, \'{E}cole Polytechnique F\'{e}d\'{e}rale de Lausanne (EPFL), CH-1015 Lausanne, Switzerland
}

\author{Jane Kim}
\affiliation{Department of Physics and Astronomy and Facility for Rare Isotope Beams, Michigan State University, East Lansing, Michigan 48824, USA}

\author{Alessandro Lovato}
\affiliation{Physics Division, Argonne National Laboratory, Argonne, Illinois 60439}
\affiliation{Computational Science Division, Argonne National Laboratory, Argonne, Illinois 60439}
\affiliation{INFN-TIFPA Trento Institute for Fundamental Physics and Applications, 38123 Trento, Italy}

\author{Giuseppe Carleo}
\affiliation{Institute of Physics, École Polytechnique Fédérale de Lausanne (EPFL), CH-1015 Lausanne, Switzerland}
\affiliation{%
Center for Quantum Science and Engineering, \'{E}cole Polytechnique F\'{e}d\'{e}rale de Lausanne (EPFL), CH-1015 Lausanne, Switzerland
}

\date{\today}

\begin{abstract}
We introduce a message-passing-neural-network-based wave function Ansatz to simulate extended, strongly interacting fermions in continuous space. Symmetry constraints, such as continuous translation symmetries, can be readily embedded in the model. We demonstrate its accuracy by simulating the ground state of the homogeneous electron gas in three spatial dimensions at different densities and system sizes.
With orders of magnitude fewer parameters than state-of-the-art neural-network wave functions, we demonstrate better or comparable ground-state energies. Reducing the parameter complexity allows scaling to $N=128$ electrons, previously inaccessible to neural-network wave functions in continuous space, enabling future work on finite-size extrapolations to the thermodynamic limit. We also show the Ansatz's capability of quantitatively representing different phases of matter.
\end{abstract}

\maketitle

\section{Introduction}
Predicting emergent physical phenomena and system properties from the \emph{ab-initio} description of the system's constituents is notoriously difficult~\cite{austin2012quantum, carlson2015quantum}. Fermionic systems, in particular, can exhibit strong correlations among the particles, leading to collective phenomena in the form of exotic phases of matter, e.g.\ superconductivity and superfluidity~\cite{RevModPhys.47.331, tinkham2004introduction}. 

In recent years, progress in numerical simulations of strongly correlated systems was triggered by the development of increasingly precise machine-learning approximation techniques. Most notably, artificial neural network (NN) architectures, in combination with Variational Monte Carlo (VMC), have shown great promise in representing ground states of quantum spin systems, especially in more than one spatial dimension~\cite{carleo2017solving, carrasquilla2020machine, PhysRevLett.121.167204, PhysRevB.100.125131, PhysRevResearch.2.023358, 10.21468/SciPostPhys.10.6.147, PhysRevX.11.031034, PhysRevX.11.041021}. 
Due to the universal approximation property of NNs, neural-network quantum states (NQS) can, in theory, accurately represent any quantum many-body state~\cite{deng2017quantum, PhysRevLett.122.065301}.
NQS have been extended to fermionic degrees of freedom in a discrete basis~\cite{diluo2019backflow, choo2020fermionic,robledo2022fermionic}, by incorporating the indistinguishability of quantum particles. More recently, advancements to ground and excited state searches for fermionic and bosonic continuous degrees of freedom with open~\cite{FermiNet, PauliNet, entwistle2023electronic} and periodic boundary conditions (PBCs)~\cite{pescia2022neural, wilson2022wave, cassella2022discovering}, have been introduced. 

The flexibility of NQS, compared to more traditional models, allows for the representation of multiple phases of matter, and even different physical systems, with a single Ansatz.
To exemplify this point, we refer to NQS studies on the ground state of molecular systems~\cite{FermiNet, PauliNet}, solutions to effective field theory Hamiltonians describing atomic nuclei~\cite{PhysRevLett.127.022502, gnech2022nuclei, lovato2022hidden}, bulk studies of fermionic and bosonic extended systems~\cite{pescia2022neural, li2022ab, cassella2022discovering, xie_mast_2022}, as well as NQS simulations of low-density neutron matter found in neutron stars~\cite{fore2022dilute}. The downside of this flexibility, especially in continuous space, is that NQS typically need a significant amount of variational parameters to reach a given accuracy. This makes optimization challenging and costly, preventing the usage of refined optimization schemes, e.g.\ second order optimization procedures~\cite{Toulouse:2007, sorella1998green}. As a result, the accessible system sizes are limited to a few tens of particles. However, studying larger system sizes is of utmost importance to estimate physical properties in the thermodynamic limit~\cite{lin2001twist, drummond2013diffusion, alavi2021benchmark, azadi2022correlation}. To remedy the situation, novel NQS architectures must be developed that significantly reduce the parameter complexity while retaining high accuracy.

This work introduces a neural-network wave function suitable for simulating strongly interacting fermionic quantum systems in continuous space with \emph{one to two} orders of magnitude fewer parameters than current state-of-the-art NQS.
The general form of the Ansatz is motivated by an analytical argument, relating the exact ground-state wave function to a many-body coordinate transformation of the electronic coordinates. It uses a permutation-equivariant message-passing architecture on a graph, inherently implementing the indistinguishability of same-species quantum particles~\cite{gilmer2017neural}. 
As an application, we study the Homogeneous Electron Gas (HEG) in three spatial dimensions without explicitly breaking any of the fundamental symmetries of the system, such as translations and spin-inversion symmetry. This allows characterizing, from first principles, the onset of Wigner crystallization at low densities. 
\section{Methods}
Throughout this work, we consider a non-relativistic Hamiltonian of identical particles with mass $m$ in $d$ spatial dimensions:  
\begin{align}
    H &= -\frac{\hbar^2}{2m}\sum_i^N \nabla_{\vec{r}_i}^2 +  V(\mathbf{X}), \label{eq:genhamiltonian}
\end{align}
where the potential and interaction energy, $V$, is assumed to be diagonal in position representation, defined by the particle coordinates $\mathbf{X} = (\vec{r}_1, ..., \vec{r}_N)$, $\vec{r}_i \in \mathbb{R}^d$. In the following, we derive an analytic functional form of the ground-state wave function and relate it to our variational Ansatz.
\subsection{Exact Backflow Transformations}
Given a suitable reference state $|\Phi_0\rangle$, as initial condition for the imaginary-time ($\tau$) evolution induced by the Hamiltonian, $
    \Phi_{\tau}(\mathbf{X}) = \langle \mathbf{X}|e^{-\tau H}|\Phi_0\rangle$,
the exact ground-state is obtained in the large imaginary-time limit:  $\lim_{\tau\rightarrow \infty}  \Phi_{\tau}(\mathbf{X}) \propto \Psi_0(\mathbf{X})$, provided $|\Phi_0\rangle$ is non-orthogonal to the exact ground state, $|\langle\Psi_0 |\Phi_0\rangle|> 0$.
For fermions, non-orthogonality implies that the wave function must be at least antisymmetric w.r.t.\ the exchange of two particles i.e.\ $\Phi_0(\mathcal{P}_{ij}(\mathbf{X}))=-\Phi_0(\mathbf{X})$ ($\mathcal{P}_{ij}$ permutes particles $i$ and $j$). 
Assuming a valid quantum reference state (twice differentiable and integrable) and finite matrix elements of the propagator~\cite{apostol1974mathematical}, we can apply the mean-value theorem to the imaginary-time evolved state, yielding:
\begin{eqnarray}
   \Phi_{\tau}(\mathbf{X}) &=&\int_{\Omega}d\mathbf{X}^{\prime}G_{\tau}(\mathbf{X},\mathbf{X}^{\prime})\Phi_{0}(\mathbf{X}^{\prime})\\
   &=&\text{Vol}(\Omega)\times G_{\tau}\left(\mathbf{X},\mathbf{Y}(\mathbf{X})\right)\Phi_{0}(\mathbf{Y}(\mathbf{X}))\label{eq::meanvalue},
\end{eqnarray}
where $\Omega$ is the (convex) integration domain for the positional degrees of freedom, and $G_{\tau}(\mathbf{X},\mathbf{X}^{\prime})=\langle\mathbf{X}|e^{-\tau H}|\mathbf{X}^{\prime}\rangle$ is the matrix element of the imaginary-time propagator, which is bounded for finite $\tau$. In Eq.~\eqref{eq::meanvalue}, we introduced the mean-value point $\mathbf{Y}(\mathbf{X})=(\vec{y}_1(\mathbf{X}), ..., \vec{y}_N(\mathbf{X})) \in \Omega$, depending parametrically on the coordinates $\mathbf{X}$.

For general Hamiltonians in the form of Eq.~\eqref{eq:genhamiltonian}, we have $G_{\tau}(\mathbf{X},\mathbf{X}^{\prime})\geq 0$, for all $\mathbf{X},\mathbf{X}^\prime$. Moreover, $G_{\tau}(\mathbf{X},\mathbf{X}^{\prime})$ is invariant under the exchange of particle coordinates: $G_{\tau}(\mathcal{P}_{ij}(\mathbf{X}),\mathcal{P}_{ij}(\mathbf{X}^{\prime}))=G_{\tau} (\mathbf{X},\mathbf{X}^{\prime})$. 
In the fermionic case, the latter implies that $\mathbf{Y}(\mathbf{X})$ must be equivariant under particle exchange, $\mathbf{Y}(\mathcal{P}_{ij}(\mathbf{X}))=\mathcal{P}_{ij}(\mathbf{Y}(\mathbf{\mathbf{X}}))$, to ensure antisymmetry of the total wave-function. Eq.~\eqref{eq::meanvalue} therefore yields the product between a permutation symmetric, positive semi-definite function $J(\mathbf{X})=G_{\tau}(\mathbf{X},\mathbf{Y}(\mathbf{X})) \times \mathrm{Vol}(\Omega)$ and a reference state computed at modified coordinates $\mathbf{Y}(\mathbf{X})$: 
\begin{align}
    \Phi_{\tau}(\mathbf{X})=J(\mathbf{X}) \times \Phi_0(\mathbf{Y}(\mathbf{X}))\label{eq::genansatz}.
\end{align}
Identification of the mean-value point $Y(X)$ with a many-body coordinate transformation gives an alternative justification for the backflow transformations~\cite{feynman1956backflow} of single-particle coordinates.
With a Slater determinant of given spin orbitals $\phi_\mu(\vec{r}_i)$ as initial state, $\Phi_0(\mathbf{X})=\det{\phi_\mu(\vec{r}_i)}/{\sqrt{N!}}$, Eq.~\eqref{eq::genansatz} is structurally related to the heuristic Jastrow-Backflow variational form~\cite{ceperley1998BF, kwon1993effects}. We remark that the symmetric contribution, $J(X)$, can be incorporated to the determinant: 
\begin{align}
    \Phi_\tau(\mathbf{X}) = \mathcal{K}\times\det{\varphi_\mu(\mathbf{y}_i(\mathbf{X}))}, \label{eq::exactbf}
\end{align}
with $\varphi_\mu(\mathbf{y}_i(\mathbf{X}))=\phi_\mu(\mathbf{y}_i(\mathbf{X}))\times \sqrt[N]{J(\mathbf{X})}$, and $\mathcal{K}$ a normalization constant.

The functional form \eqref{eq::genansatz},\eqref{eq::exactbf} is exact, provided that the symmetric factor $J(\mathbf{X})$ and the mean-value coordinates $\mathbf{Y}(\mathbf{X})$ satisfy Eq.~(\ref{eq::meanvalue}), and the reference state is not orthogonal to the exact ground state.
An approximate but explicit form for the coordinate transformation $\mathbf{Y}(\mathbf{X})$ can be obtained by repeatedly applying the imaginary-time propagator to the reference state in the limit of small $\tau$. This process gives rise to the iterative backflow transformation, as introduced in Ref.~\cite{taddei2015iterative, markus2018backflow}. 

\subsection{Message-Passing Neural Quantum States}
\begin{figure}[t!]
  \includegraphics[width=\columnwidth]{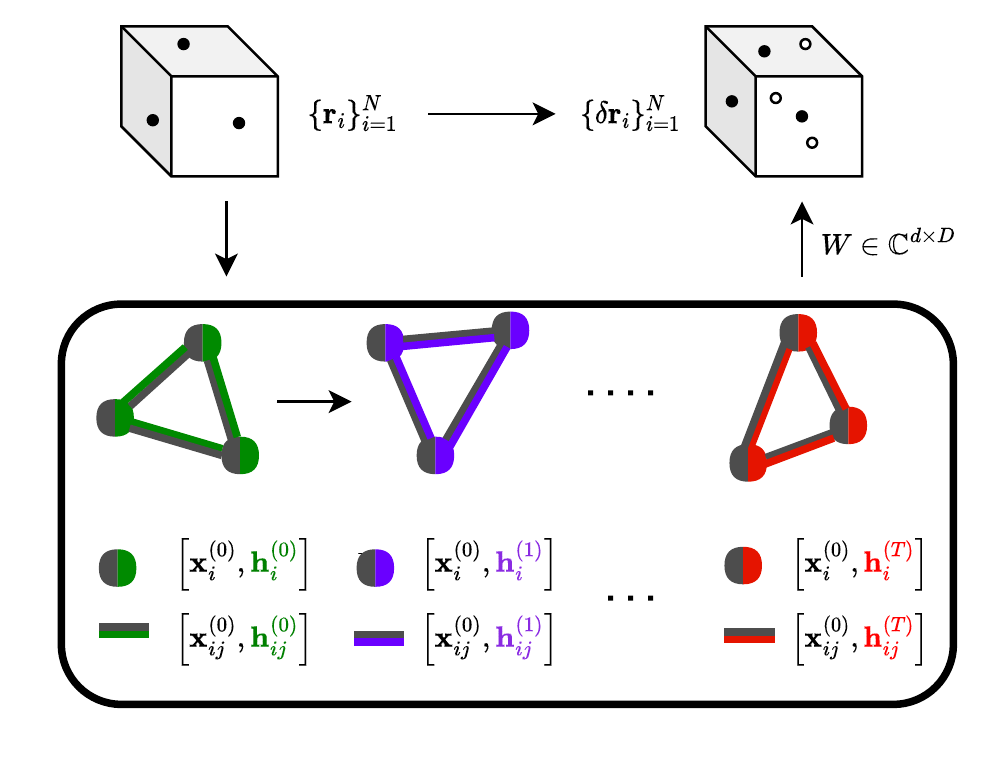}
  \caption{\label{fig:MPNN} Illustration of backflow transformation via MPNN. Single-particle coordinates $\vec{r}_i \in \mathbb{R}^d$ (black dots, top left) are mapped to quasi-particle coordinates $\delta\vec{r}_i \in \mathbb{C}^d$(black/white dots for real/imaginary part, top right). From an initial graph consisting of an initial feature vector (dark grey) and a hidden state (green), a new graph, consisting of the initial features and an updated hidden state (indicated by different coloring), is constructed via messages. The final node states are linearly transformed back to $\delta\vec{r}_i = W\cdot\mathbf{g}_i^{(T)}$, containing information about all particles ($D$ is the dimension of the last graph's nodes).}
\end{figure}

Motivated by Eq.~\eqref{eq::exactbf}, we use single-particle orbitals, $\{\phi_\mu\}_{\mu=1}^N$, evaluated at many-body backflow coordinates, $\mathbf{Y}(\mathbf{X})$, to construct the variational Ansatz. The backflow transformation is parameterized with permutation-equivariant message-passing NNs (MPNN)~\cite{gilmer2017neural} (see Fig.~\ref{fig:MPNN}), hence we name it Message-Passing Neural Quantum State (MP-NQS).
In the MPNN, an all-to-all connected graph, encoding effective particle positions (nodes) and their interactions (edges):
\begin{align}
    \mathbf{g}_i^{(t)} &= [\mathbf{x}_i^{(0)}, \mathbf{h}_i^{(t)}] \in \mathbb{R}^{D_1} \\
    \mathbf{g}_{ij}^{(t)} &= [\mathbf{x}_{ij}^{(0)}, \mathbf{h}_{ij}^{(t)}] \in \mathbb{R}^{D_2},
\end{align} 
is updated iteratively. Here, we have introduced a discrete step index $t\geq 0$, initial feature vectors $\mathbf{x}_i^{(0)}\in\mathbb{R}^{D_1^{(0)}}$, $\mathbf{x}_{ij}^{(0)}\in\mathbb{R}^{D_2^{(0)}}$, and we denote concatenation with $[\cdot, \cdot]$. The iterative equations above also contain auxiliary variables known as hidden states ($\mathbf{h}_{i}^{(t)} \in \mathbb{R}^{D^h_1}$, $\mathbf{h}_{ij}^{(t)} \in \mathbb{R}^{D^h_2}$) with suitably chosen \emph{feature} dimensions such that $D_1^{h} + D_1^{(0)} = D_1$, and $D_2^{h} + D_2^{(0)} = D_2$. 
Construction of the initial feature vectors ($\mathbf{x}_i^{(0)}$, $\mathbf{x}_{ij}^{(0)}$) is system dependent and will be discussed in detail later.

The hidden states are initialized, independently of their index, with vectors of learnable parameters. They are updated using permutation-equivariant \emph{messages}, obtained from a variant of the attention mechanism~\cite{vaswani2017attention}, we dub here \textit{particle attention}.
Specifically, the messages are given by weighted transformations of the edges $\mathbf{g}_{ij}^{(t)}$: $\mathbf{m}_{ij}^{(t+1)} = \boldsymbol{\omega}^{(t)}_{ij}(\mathbf{g}_{ij}^{(t)}) \odot \boldsymbol{\phi}(\mathbf{g}_{ij}^{(t)})$, where $\odot$ represents element-wise multiplication along the feature dimension, and $\boldsymbol{\omega}^{(t)}_{ij} \in \mathbb{R}^{D_2}$ are weight vectors. The weights are obtained using \emph{query}/\emph{key} matrices given by 
$\mathbf{Q}_{ij}^{(t)} = W_{Q} \cdot \mathbf{g}_{ij}^{(t)}$ and $\mathbf{K}_{ij}^{(t)} = W_{K} \cdot \mathbf{g}_{ij}^{(t)}$, with weight matrices $W_{Q}, W_{K} \in \mathbb{R}^{D_2 \times D_2}$.
Applying an element-wise GELU non-linearity~\cite{hendrycks2016gaussian} to the overlap between queries and keys along the \emph{particle dimension} (as opposed to the feature dimension~\cite{vaswani2017attention, von2022self}), results in permutation-equivariant weights 
\begin{align}
\boldsymbol{\omega}^{(t)}_{ij}&=\mathrm{GELU}\left(\sum_l \mathbf{Q}_{il}^{(t)} \mathbf{K}_{lj}^{(t)}\right).
\end{align}
The particle attention compares environments of particles $i$ and $j$, and effectively increases the order of correlations that can be embedded in a single iteration of the network. This is crucial to reduce the total number of network iterations (parameters) and capture many-body effects. 
The hidden states are updated using the current graph and messages:
\begin{align}
    \mathbf{h}_i^{(t+1)} &= \boldsymbol{f}\left(\mathbf{g}_i^{(t)}, \sum_{j\neq i}\mathbf{m}_{ij}^{(t+1)} \right) \label{eq::singlestream} \\
    \mathbf{h}_{ij}^{(t+1)} &= \boldsymbol{\Tilde{f}}\left(\mathbf{g}_{ij}^{(t)}, \mathbf{m}_{ij}^{(t+1)} \right) \label{eq::twostream}
\end{align}
The functions $\boldsymbol{\phi}$, $\boldsymbol{f}$, and $\boldsymbol{\Tilde{f}}$ are parameterized by Multilayer Perceptrons (MLPs).
The updated graph then has the same structure as the former: $\mathbf{g}_i^{(t+1)} = [\mathbf{x}_i^{(0)}, \mathbf{h}_i^{(t+1)}] \in \mathbb{R}^{D_1}$, $\mathbf{g}_{ij}^{(t+1)} = [\mathbf{x}_{ij}^{(0)}, \mathbf{h}_{ij}^{(t+1)}] \in \mathbb{R}^{D_2}$.
Inclusion of the initial inputs, referred to as a ``skip connection" in ML literature~\cite{he2016deep}, mitigates the vanishing gradient problem and allows a more efficient capture of correlations.

The final backflow coordinates are constructed as $\vec{y}_i(\mathbf{X})  = \vec{r}_i + \delta\vec{r}_i(\mathbf{X})$, where the displacements, $\delta\vec{r}_i \left(\mathbf{X} \right)$, are obtained via a linear transformation of the final node states to $d$ dimensions, i.e.\ $\delta\vec{r}_i \left(\mathbf{X}\right) = W \cdot \mathbf{g}_i^{(T)}$ with $W \in \mathbb{C}^{d \cross D_1}$. The complex-valued backflow transformation allows changing the degree of localization, determined by the chosen single-particle orbitals, and representing complex-valued wave functions in general.

Following \eqref{eq::exactbf}, we further augment the orbitals with a permutation-invariant factor $J$, of the form 
\begin{align}
    J(\mathbf{Y}, \mu) = \sum_i j(\mathbf{y}_i, \mu)
\end{align}
where $j$ is parameterized with a MLP, and $\mu$ denotes the quantum numbers of the orbitals, yielding:
\begin{align}
    \Psi(\mathbf{X}) = \det{\varphi_\mu(\mathbf{y}_i(\mathbf{X})) }, 
\end{align}
with $\varphi_\mu(\vec{y}_i) = \exp\left[J(\mathbf{Y}, \mu)\right]\times \phi_\mu(\vec{y}_i)$.
\subsection{Electron Gas}
We now study the case of the homogeneous electron gas (HEG) in $d=3$ spatial dimensions, a prototypical model for the electronic structure in solids. It includes Coulomb interactions among the solids' electrons while treating its positively charged ions as uniform, static, positive background~\cite{pines1963elementary}. Despite this simplification, the HEG exhibits different phases of matter and captures properties of real solids, particularly of Alkali metals. 
The Hamiltonian (in units of Hartree), for a system of $N$ electrons with uniform density $n=\frac{N}{V}$, is given by:
\begin{align}
    H &= -\frac{1}{2r_s^2}\sum_i^N \nabla_{\vec{r}_i}^2 + \frac{1}{r_s}\sum_{i<j}^N \frac{1}{\norm{\vec{r}_i - \vec{r}_j}} + \mathrm{const.} \label{eq:hamiltonian}
\end{align}
where we introduced the Wigner-Seitz radius $r_s = \sqrt[3]{3/(4 \pi n)}$, and a constant arising from the electron-background interaction~\cite{ceperley1998BF}. 
The conditionally convergent series of pairwise Coulomb interactions is evaluated using the Ewald summation technique, as is standard for extended systems in QMC~\cite{ewald1921berechnung, foulkes1996ewald, toukmaji1996ewald}. We will assume a fixed spin-polarization with $N=N_\uparrow + N_\downarrow$, where $N_{\uparrow/\downarrow}$ denotes the number of up/down spins. Additionally, $s_i \in \{\uparrow, \downarrow \}$ denotes the spin of the $i$-th electron. We equip the cubic simulation cell of side length $L$ with periodic boundary conditions (PBCs) in all spatial directions to access the bulk of the system.

As in \eqref{eq::exactbf}, we use a single Slater determinant as a reference state, $\Phi_0(\mathbf{X})$.
We expect a liquid-crystal phase transition for the HEG, as a function of the density $n$. The dominating kinetic energy in Eq.~\eqref{eq:hamiltonian} ($\sim 1/r_s^2$) for large $n$ leads to the well-known Fermi liquid behavior. For small $n$, the potential energy ($\sim1/r_s$) dominates and enforces a crystalline BCC structure among the electrons, known as Wigner crystal~\cite{PhysRev.46.1002}. 
Plane-wave orbitals are a natural and physically-motivated choice to model the liquid, translation invariant phase: $\phi_\vec{k}(\vec{r}) = \exp\left[i\vec{k} \cdot\vec{r}\right]$ with $\vec{k} = \frac{2\pi}{L} \vec{n}$ where $\vec{n}\in \mathbb{Z}^d$. These orbitals allow modeling the translation invariant system at fixed total momentum $\vec{k}_{\textrm{tot}} = \sum_{i=1}^{N} \vec{k}_i$. To account for the spin $s$ of a particle located at $\mathbf{r}$, we use spin-orbitals $\phi_\mu(\vec{r}, s) = \phi_{\vec{k}_\mu}(\vec{r})\delta_{s_\mu, s}$, where each spin-orbital is characterized by the quantum numbers $\mu = (\vec{k}_\mu, s_\mu)$. This choice of spin-orbitals lets the determinant factorize into a product of determinants of up and down spin orbitals.
To study the Wigner crystal phase, we employ localized Gaussian orbitals centered around the BCC lattice sites $\mathbf{R}_\mu$: $\phi_\mu(\mathbf{r}, s) = \sum_{\mathbf{R}_n} \exp\left[-\alpha(\mathbf{r}-\mathbf{R}_\mu +\mathbf{R}_n)\right])\delta_{s_\mu, s}$, where $\alpha$ is a variational parameter, $\mu = (\mathbf{R}_\mu, s_\mu)$, and we sum over simulation cell lattice vectors $\mathbf{R}_n$ to ensure periodicity.
For all densities we use a simple cubic simulation cell to not bias towards any one of the phases.

We further specialize the MP-NQS to the HEG by defining initial feature vectors. Respecting the spin inversion and translation symmetries of the HEG requires us to ignore single-particle positions and spins. We, therefore, initialize the nodes to a learnable embedding vector $\mathbf{e}\in \mathbb{R}^{D_1-D_1^h}$, that does not depend on the particle index $i$. For the edge features, we use the translation invariant particle-distances $\vec{r}_{ij} = \vec{r}_i - \vec{r}_j$ and their norm. Same- and opposite-spin pairs are distinguished using products of the form $s_i\cdot s_j = \pm 1$ in the edge features. Overall, we obtain the following initial features: 
\begin{align}
    \mathbf{x}_i^{(0)} = \mathbf{e}\mathrm{,} \qquad \mathbf{x}_{ij}^{(0)} = [\mathbf{r}_{ij}, \norm{\mathbf{r}_{ij}}, s_i\cdot s_j].
\end{align}
Notice that this choice preserves the spin quantum  number of each particle.

The PBCs of the simulation box are incorporated by mapping the components of vectors $\vec{r} \in \mathbb{R}^d$ (where $\vec{r} = \vec{r}_i$ or $\vec{r}=\vec{r}_{ij}$) to a Fourier basis $\vec{r} \mapsto \left[\sin(\frac{2\pi}{L}\vec{r}), \cos(\frac{2\pi}{L}\vec{r})\right]\in \mathbb{R}^{2d}$, and the norm of the distance between two particles, $\norm{\vec{r}_{ij}}$, to a periodic surrogate $\norm{\vec{r}_{ij}} \mapsto \norm{\sin(\frac{\pi}{L}\vec{r}_{ij})}$, as in Ref.~\cite{pescia2022neural}. In summary, our Ansatz exhibits translation and spin-inversion invariance and maintains a fixed total momentum, $\vec{k}_{\textrm{tot}}$. Its number of parameters is independent of system size (here $\sim 19000$) and, using Stochastic Reconfiguration (SR)~\cite{sorella1998green}, only $\mathcal{O}(10^3)$ optimization steps are needed to reach convergence. A comparison to other NQS approaches is given in the Supplemental Material.

\section{Results}
We study the fully spin-polarized and unpolarized HEG in different density regimes $r_s \in [1,200]$ and up to system sizes of $N=128$ electrons. We compare our ground-state energies against state-of-the-art NQS architectures -- FermiNet~\cite{cassella2022discovering} and WAPNet~\cite{wilson2022wave} -- for small system sizes $N\in\{14,19\}$ and against \emph{Diffusion Monte Carlo} (DMC) with backflow (BF-DMC) ~\cite{PhysRevE.74.066701,PhysRevB.105.245135} for $N=54$ electrons. Additional benchmarks w.r.t.\ common quantum chemistry methods, including transcorrelation augmented full configuration interaction method (FCI) and distinguishable clusters with doubles (DCD) method~\cite{alavi2021benchmark}, are included in the Supplementary Material.
The effect of the backflow transformation on the nodal surface is studied by comparing to fixed-node DMC (FN-DMC) results. We use an energy of $1.5$ mHa per particle (chemical accuracy) to assess the significance of energy differences between the different methods. An overview of all results and benchmarks for the various system sizes and densities is provided in the Supplemental Material.
\subsection{Small Systems.--}
\begin{figure}[t!]
  \includegraphics[width=\columnwidth]{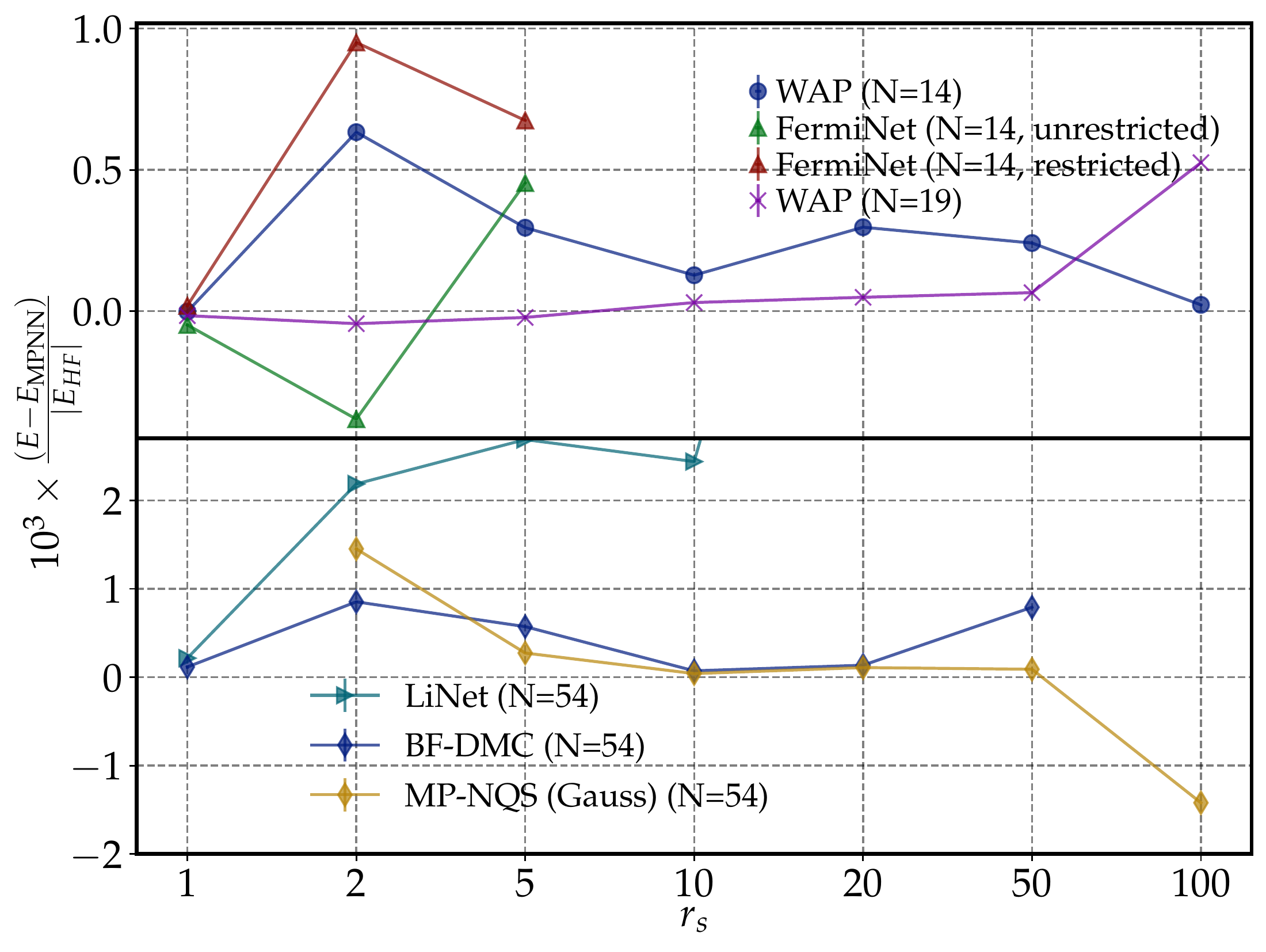}
  \caption{\label{fig::energies}Comparison of ground-state energies obtained with different methods for various densities, polarizations, and system sizes, in mHa per particle. Each line shows the difference in energy of the respective method to the MP-NQS energy with plane-wave orbitals $E_{\mathrm{MPNN}}$ (values above zero are worse than the MP-NQS baseline). (Top) $N=14,19$, (Bottom):  $N=54$ particles. Error bars are too small to be visible for most densities. The corresponding numerical data can be found in the Supplemental Material.}
\end{figure}

The best available results for $N=14$ are obtained with the FCI method, compared to which we find an energy difference of less than $1.5$ mHa per particle. State-of-the-art NQS architectures perform comparably to the MP-NQS: The unrestricted FermiNet performs slightly better ($\mathcal{O}\left(10^{-5}\right)$ Ha/N) than both MP-NQS and WAPNet for $r_s \leq 2$, while the MP-NQS and WAPNet improve over this version of FermiNet for $r_s=5$. The restricted FermiNet yields worse ground-state energies than the MP-NQS over all probed densities~\cite{cassella2022discovering} (see Fig.~\ref{fig::energies}). It should be noted that the energy differences for this system are of the order of machine precision.
For $r_s \geq 5$, we find slightly better performance than WAPNet for all of the reported densities. All differences lie within a range of $1.5$ mHa per particle.
Similarly, for $N=19$, the MP-NQS obtains slightly higher energies than WAPNet for large densities ($r_s\leq 5)$ and marginally lower ones at smaller densities ($r_s\geq 5$) with differences lower than $1.5$ mHa per particle.

\subsection{Large Systems.--}
\begin{figure}[t!]
  \includegraphics[width=\columnwidth]{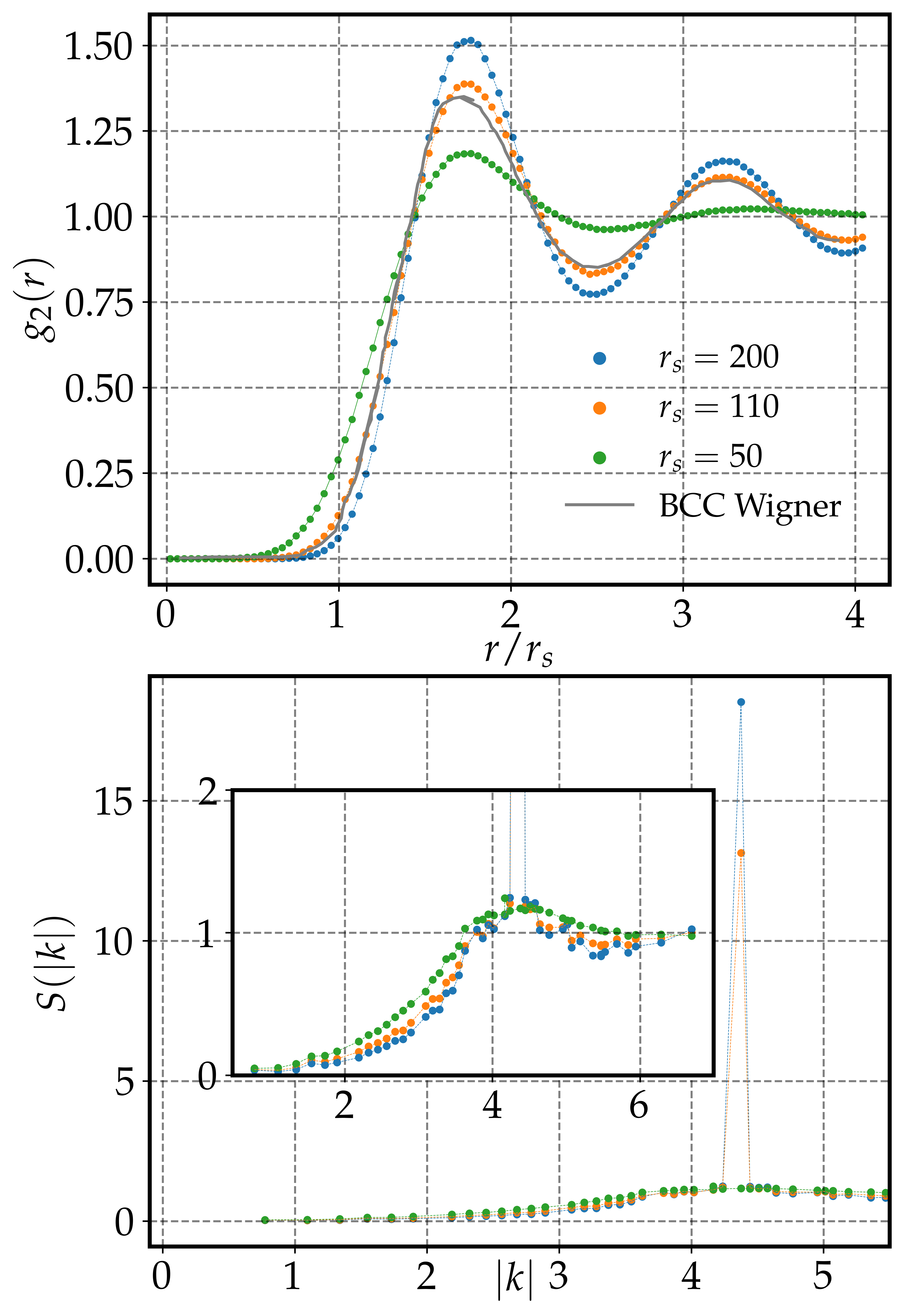}
  \caption{\label{fig::g2N128} Spin-averaged radial distribution function (top) and corresponding structure factor (bottom) for the HEG system with $N=128$ electrons at $r_s=50,110,200$. For $r_s=50$ we used plane-waves as reference state while for $r_s=110,200$ Gaussian orbitals centered at BCC lattice sites were used. Error bars are smaller than the symbols. The crystal and liquid benchmarks are obtained from \cite{drummond2004diffusion} for $r_s=110$.}
\end{figure}

For $N=54$ particles, accurate results are obtained with the FCI method~\cite{alavi2021benchmark}. At this system size, the energy differences per particle are smaller than for $N=14$, suggesting size-consistency of our Ansatz, provided that the FCI method is size consistent as well. When compared to purely variational methods, we obtain significantly better ground-state energies than FN-DMC and BF-DMC, especially at high densities. This is in stark contrast to the (FermiNet-based) architecture of~\cite{li2022ab} (dubbed LiNet in the following), which does not improve upon BF-DMC energies over the whole density regime (see Fig.~\ref{fig::energies}, bottom).
At $r_s \geq 50$, we observe improved results using the MP-NQS with Gaussian orbitals compared to plane-waves (see Fig.~\ref{fig::energies}, bottom panel, yellow line). This strongly suggests a transition from a de-localized Fermi liquid to a localized crystalline phase as expected from previous studies~\cite{drummond2004diffusion,azadi2022correlation}. To analyze finite size effects, we also simulate a larger system of $N=128$ electrons at $r_s=50, 110, 200$, and confirm that Gaussian orbitals lead to lower ground-state energies, compared to the plane-waves for $r_s > 50$ (see Supplementary Material). Furthermore, for $r_s=110,200$ the crystalline character of the variational state can be clearly seen in the radial distribution functions and corresponding structure factors displayed in Fig.~\ref{fig::g2N128}. The prominent peak in the structure factor and the pronounced density fluctuations in the radial distribution function up to the maximum distance of $L/2$, indicate the crystalline nature of the represented state. Note that these are absent for $r_s=50$, suggesting a fluid state.

\section{Conclusions.--} We have introduced MP-NQS, a novel NQS architecture that leverages MPNNs to build highly expressive backflow coordinates. We demonstrate its power on the HEG system, reducing the number of parameters by orders of magnitudes compared to state-of-the-art NQS in continuous space while reaching at par or better accuracy. We also show improvement upon state-of-the-art BF-DMC results on large systems.
The favorable scaling allows us to accurately simulate large periodic electronic systems, previously inaccessible to state-of-the-art NQS models. We increase the available system sizes from $N=27$ and $N=54$ electrons in periodic systems~\cite{li2022ab,cassella2022discovering,wilson2022wave} to $N=128$ electrons in this work. Hence we open the door to extrapolation methods to the thermodynamic limit for extended systems. Furthermore, we reproduce the liquid-crystal phase transition of the HEG around $r_s=100$, matching previous predictions on the transition density~\cite{ceperley1980heg,drummond2004diffusion, azadi2022correlation}, showing the MP-NQS capability to describe different phases of matter quantitatively better than previous studies of the HEG based on neural quantum states~\cite{cassella2022discovering}.
In addition to the numerical results, we also introduced an analytical argument, justifying commonly adopted backflow transformations. Our argument shows that a backflow transformation over a reference state is sufficient to obtain the exact ground-state wave function. It will be of particular interest to characterize the geometrical properties of these transformations and understand in what cases neural-network parameterizations can efficiently describe them.   

\begin{acknowledgments} 
We acknowledge stimulating discussions with Markus Holzmann, and Francesco Pederiva. 
The numerical tools used in this work are based on the open-source software NetKet\cite{carleo_netket:_2019, hafner2021mpi4jax, vicentini2022netket} version 3. The present research is supported by the Swiss National Science Foundation under Grant No. 200021\_200336, by the NCCR MARVEL, a National Centre of Competence in Research, funded by the Swiss National Science Foundation (grant number 205602), Microsoft Research, by the U.S. Department of Energy, Office of Science, Office of Nuclear Physics, under contracts DE-AC02-06CH11357 (A.L), by the 2020 DOE Early Career Award program (A.L.), the NUCLEI SciDAC program (A.L.), and Argonne LDRD awards (A. L.). 
J.~K. is supported by the U.S. Department of Energy, Office of Science, office of Nuclear Physics under grant No. DE-SC0021152 and U.S. National Science Foundation Grants No. PHY-1404159 and PHY-2013047.
\end{acknowledgments}
\bibliography{biblio}
\appendix
\section{Optimization \label{Optimization}}
From the Rayleigh-Ritz principle, we obtain a lower bound on the expectation value of the Hamiltonian $\mel{\Psi}{H}{\Psi}/\braket{\Psi}{\Psi} \equiv E[\Psi] \geq E_0$. We use the energy expectation value with respect to our variational Ansatz as a cost function to gauge the proximity of the variational state to the ground state of the Hamiltonian.
Formally the ground state is given by:
\begin{equation}
\Psi_0 = \argmin\limits_{\Psi} E[\Psi]\, .
\end{equation}
where $H\ket{\Psi_0} = E_0\ket{\Psi_0}$.

Since the exact expected value of the Hamiltonian requires us to analytically solve a high-dimensional integral, which in general is not feasible, we resort to Monte Carlo sampling and integration. The energy expectation is evaluated as the average over a set of local energy 
$E_{\mathrm{loc}}(\mathbf{X}) = \mel{\mathbf{X}}{H}{\Psi}/\bra{\mathbf{X}}\ket{\Psi}$, where the samples $\mathbf{X}$ are obtained from the probability distribution $|\Psi(\mathbf{X})|^2/\bra{\Psi}\ket{\Psi}$ using the Metropolis-Hastings algorithm.

To update the variational parameters in our Ansatz such that it exhibits progressively lower energy expectation values, we apply the stochastic reconfiguration (SR) algorithm~\cite{sorella1998green}, which can be shown to be equivalent to imaginary-time evolution on the variational manifold and is related to the Natural Gradient descent method.
The update rule for the variational parameters is given by $\delta \boldsymbol{\theta} = -\eta S^{-1}\mathbf{F}$, where $\mathbf{F}$ is the force vector consisting of the log-derivatives of the energy w.r.t.\ the variational parameters $\boldsymbol{\theta}$ given by
\begin{align}
    F_i=2\left( \frac{\bra{\partial_i\Psi} H \ket{\Psi}}{\braket{\Psi}{\Psi}}-E[\Psi] \frac{\braket{\partial_i\Psi}{\Psi}}{\braket{\Psi}{\Psi}}\right),
\end{align}
and the so-called quantum geometric tensor $S$ is
\begin{align}
    S_{ij}=\frac{\bra{\partial_i\Psi}\ket{\partial_j\Psi}}{\braket{\Psi}{\Psi}}-\frac{\bra{\partial_i \Psi}\ket{\Psi}\bra{\Psi}\ket{\partial_{j}\Psi}}{\braket{\Psi}{\Psi}\braket{\Psi}{\Psi}},
\end{align}
and $\eta$ is the learning rate~\cite{neuscamman2012optimizing}. To regularize the computation of $S^{-1}$, we apply a small diagonal shift such that we actually compute $(S+\epsilon \mathbb{I})^{-1}$. For all experiments, we used $\epsilon=10^{-4}$. The learning rate was chosen dependent on the density as $\eta=\{0.05,0.05,0.05,0.1,0.1,0.5,1,2.5\}$ for $r_s = \{1,2,5,10,20,50,100,110\}$.\\

\section{Message-Passing Neural Networks}
In the following, we provide an overview on the general structure of message-passing neural networks~\cite{gilmer2017neural,batatia2022mace} (MPNNs) and relate it to the MP-NQS.
First the relevant data has to be encoded into a graph structure, where the nodes of the graph $\mathbf{g}_i^{(t)}=[\mathbf{x}_i^{(0)}, \mathbf{h}_i^{(t)}]$ describes single-body information and the edges between nodes $\mathbf{g}_{ij}^{(t)}=[\mathbf{x}_{ij}^{(0)}, \mathbf{h}_{ij}^{(t)}]$ contain relational information about the connected nodes. The graph $(\mathbf{g}_i^{(t)}, \mathbf{g}_{ij}^{(t)})$ then gets successively updated over $T$ iterations, using a two-step message-passing procedure. In the first step, the messages are typically constructed for all the nodes of the graphs, but not the edges:
\begin{align}
    \mathbf{m}_i^{(t+1)} = \sum_{j\in\mathcal{N}(i)} \mathbf{M}^{(t)}(\mathbf{g}_i^{(t)}, \mathbf{g}_j^{(t)}, \mathbf{g}_{ij}^{(t)})
\end{align}
where $\mathbf{M}^{(t)}$ is a learnable function and the sum runs over all nodes in the neighborhood of node $i$, denoted by $\mathcal{N}(i)$. The second step then consists of updating the graph nodes, using the constructed messages:
\begin{align}
    \mathbf{g}_i^{(t+1)} = \mathbf{U}^{(t)}(\mathbf{g}_i^{(t)}, \mathbf{m}_i^{(t+1)})
\end{align}
where $ \mathbf{U}^{(t)}$ is again a learnable function.
In the MP-NQS, the message construction is done via the particle attention mechanism. Additionally the explicit dependence on the nodes of the graph is removed and an update of the graph edges is added:
\begin{align}
    \mathbf{m}_{ij}^{(t+1)} &= \mathbf{M}^{(t)}(\mathbf{g}_{ij}^{(t)}) \\
    \mathbf{m}_i^{(t+1)} &= \sum_{j\neq i} \mathbf{m}_{ij}^{(t+1)} \label{eq::messages}
\end{align}
where $\mathcal{N}(i)$ is chosen to be all other particles, to retain permutation equivariance. 
The nodes and edges of the graph are then updated as follows:
\begin{align}
    \mathbf{g}_i^{(t+1)} &= \mathbf{U}_1(\mathbf{g}_i^{(t)}, \mathbf{m}_i^{(t+1)}) \\
    \mathbf{g}_{ij}^{(t+1)} &= \mathbf{U}_2(\mathbf{g}_{ij}^{(t)}, \mathbf{m}_{ij}^{(t+1)})
\end{align}
where we introduced another two learnable functions $\mathbf{U}_1,\mathbf{U}_2$.

In traditional MPNN applications, the intermediate and final graphs' nodes are processed, in a so-called \textit{readout phase}, to predict a scalar quantity, e.g. the energy of an atom in a specific molecular configuration:
\begin{align}
    E_i = \sum_t R^{(t)}(\mathbf{g}_i^{(t)})
\end{align}
where $R^{(t)}$ is again a learnable function. In the MP-NQS, we replace this step with a transformation of the final graph's nodes to the physical configuration space of the particles:
\begin{align}
   \mathbf{{} \delta r}_i = \mathbf{W} \cdot \mathbf{g}_i^{(T)} \in \mathbb{R}^d
\end{align}

\section{Attention Mechanism}
In this section, we summarize the original self-attention mechanism introduced in \cite{vaswani2017attention} and provide a comparison to the particle attention mechanism introduced in this work.
The attention mechanism builds on the query/key/value concept, notably recognized in search algorithms. Given a query vector (e.g. search phrase), it gets compared to a set of key vectors associated to candidates of a database (e.g. words on websites). The comparison of the query and key provides a weight determining the relevance of a corresponding value (e.g. website) given the query. Given (single-body) input data $\{\mathbf{x}_1,...,\mathbf{x}_N\},\quad \mathbf{x}_i \in \mathbb{R}^D$, keys/queries/values are formally computed using learnable weight matrices $W_i^{(Q)}\in \mathbb{R}^{D\times D_k}$, $W_i^{(K)}\in \mathbb{R}^{D\times D_k}$, $W_i^{(V)}\in \mathbb{R}^{D\times D_v}$:
\begin{align}
    \mathbf{Q}_i = W_i^{(Q)} \mathbf{x}_i,\quad \mathbf{K}_i = W_i^{(K)} \mathbf{x}_i,\quad \mathbf{V}_i = W_i^{(V)} \mathbf{x}_i
\end{align}
To obtain permutation-invariant self-attention from the above, the weight matrices cannot depend on the index $i$ but are the same independent of the input data:
\begin{align}
    \mathbf{Q}_i = W^{(Q)} \mathbf{x}_i,\quad \mathbf{K}_i = W^{(K)} \mathbf{x}_i,\quad \mathbf{V}_i = W^{(V)} \mathbf{x}_i
\end{align}
Comparing query and key vectors is commonly performed using the (scaled) dot-product along their feature dimension (with dimension $D_k$), such that the self-attention map can be written as:
\begin{align}
    \mathrm{\mathbf{Att}}(\mathbf{Q},\mathbf{K},\mathbf{V}) = \mathrm{softmax}(\frac{\mathbf{Q}^T\mathbf{K}}{\sqrt{d}}) \mathbf{V} \in \mathbb{R}^{N\times D_v}
\end{align}
where $\mathbf{Q},\mathbf{K},\mathbf{V}$ are matrices of stacked queries, keys and values, respectively ($\mathbf{Q}=(\mathbf{Q}_1,...,\mathbf{Q}_N)\in \mathbb{R}^{N\times D_k}$). The above can also be written as a weighted sum of value vectors, where the weights are given by the inner product between queries and keys.

In the MP-NQS we develop and extend the idea of self-attention to two-body input data $\{\mathbf{x}_{11},...,\mathbf{x}_{1N},...\mathbf{x}_{NN}\},\quad \mathbf{x}_{ij} \in \mathbb{R}^D$. As in self-attention, we utilize query/key weight matrices to compute query/key pairs:
\begin{align}
    \mathbf{Q}_{ij} &= W^{(Q)} \mathbf{x}_{ij} \in \mathbb{R}^{D_k }\\ 
    \mathbf{K}_{ij} &= W^{(K)} \mathbf{x}_{ij}\in \mathbb{R}^{D_k} \\
\end{align}
The comparison between queries and keys is again performed by taking their inner product but importantly, along the \emph{particle dimension} rather than the feature dimension. Additionally, we replace the softmax-activation with a GELU-activation:
\begin{align}
\boldsymbol{\omega}_{ij}&=\mathrm{GELU}\left(\sum_l \mathbf{Q}_{il} \mathbf{K}_{lj}\right).
\end{align}
where we introduced the weights $\boldsymbol{\omega}_{ij}$. Instead of using a linear map for the values (as is done for the queries and keys), we use a full MLP to compute values, $\mathbf{V}_{ij} = \boldsymbol{\phi}(\mathbf{x}_{ij})$ from the input data. We obtain the following attention mechanism, dubbed particle attention:
\begin{align}
    \mathrm{\mathbf{PAtt}}(\mathbf{Q},\mathbf{K},\mathbf{V}) = \boldsymbol{\omega}_{ij} \odot \mathbf{V}_{ij} \in \mathbb{R}^{N\times N \times D_v}
\end{align}
where $\odot$ denotes element-wise multiplication along the feature dimension. Using the above as messages for the MPNN, \eqref{eq::messages}, results in permutation-equivariant, weighted sums of edge features:
\begin{align}
    \mathbf{m}_{ij}^{(t)} &= \mathrm{\mathbf{PAtt}}(\mathbf{Q},\mathbf{K},\mathbf{V}) = \boldsymbol{\omega}_{ij} \odot \mathbf{V}_{ij}\\
    \mathbf{m}_i^{(t)} &= \sum_{j\neq i} \boldsymbol{\omega}_{ij} \odot \mathbf{V}_{ij}
\end{align}

\section{MP-NQS}
\begin{figure}
  \includegraphics[width=\columnwidth]{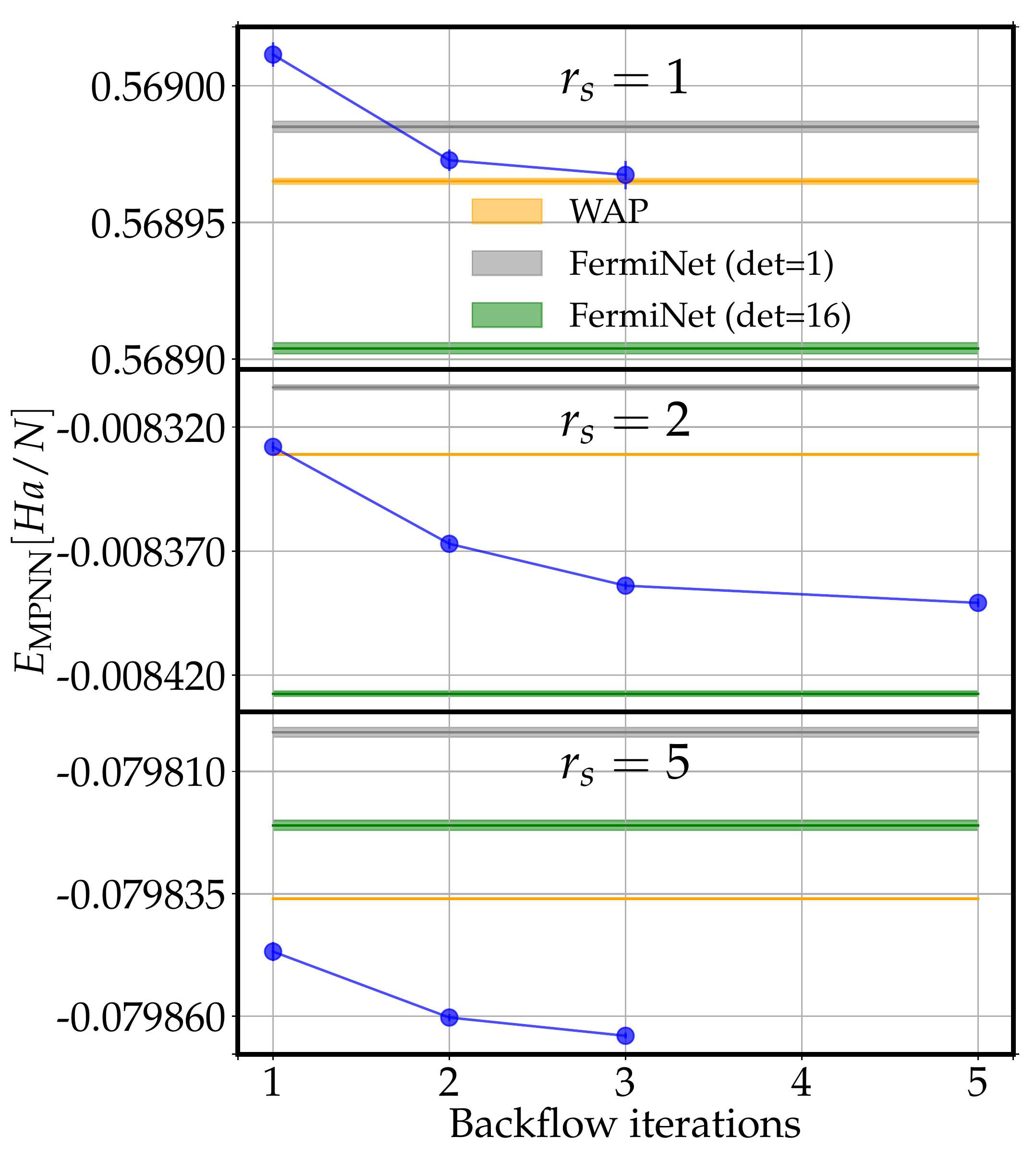}
  \caption{\label{fig::energyvsdepth} Energy obtained with our variational Ansatz for $N=14$ particles at the densities $r_s=1,2,5$ as a function of the number of message-passing backflow iterations. We compare to WAPNet and FermiNet with a single Slater determinant as well as FermiNet with 16 determinants.}
\end{figure}

We show in Figure \ref{fig::energyvsdepth}, for a system of $N=14$ particles at the highest densities $r_s=1,2,5$, that increasing the number of backflow iterations (and therefore the number of variational parameters) systematically improves the accuracy of our Ansatz. At $r_s=1$ we score slightly worse than WAPNet for a single backflow iteration, while reaching the same accuracy for three backflow iterations. Similarly, for $r_s=2$ we obtain a slightly higher energy than WAPNet for a single backflow iteration, but we surpass WAPNet's performance after two backflow iterations. For $r_s=5$ we obtain more precise results than WAPNet for a single backflow iteration. Upon comparison to FermiNet, our results demonstrate that we consistently outperform the unrestricted, single Slater determinant version of FermiNet at these densities. Interestingly, for lower densities ($r_s=5$), our Ansatz even outperforms the unrestricted FermiNet with 16 determinants with a single message-passing iteration.
\section{Comparison to other NQS architectures}\label{sec:comparison_nqs}
The most thoroughly investigated NQS architectures for continuous space are FermiNet and PauliNet~\cite{FermiNet, PauliNet}. Both have been shown to provide good approximations to the ground-state energies of molecular systems. FermiNet has additionally been applied to the HEG system with very accurate results~\cite{cassella2022discovering}. Although FermiNet shows slightly better results on molecular systems than PauliNet, differences in the architectures allow PauliNet to reach comparable energies with considerably less parameters than FermiNet. Here we compare our MP-NQS to these two Ansaetze and highlight similarities and differences.

Both FermiNet and PauliNet construct the variational wave function Ansatz by using the antisymmetry of the Slater determinant. PauliNet uses physically motivated orbital functions as input to the Slater determinant and alters these by an orbital-aware multiplicative symmetric factor. Additionally, an overall Jastrow factor is used to increase the expressiveness of the Ansatz. FermiNet imposes no physical knowledge on the orbitals (up to an overall envelope function) and does not need an overall Jastrow factor. To construct the multiplicative factor in PauliNet and the orbitals in FermiNet, the number of parameters scales linearly with the system size. In addition, both Ansaetze construct multiple Slater determinants to increase their accuracy, thereby multiplying the number of parameters by a constant factor. Both Ansaetze treat the different spin species by utilizing distinct parameter sets for the different possible spin projections.

Similar to PauliNet, we use physically motivated orbitals which are evaluated at backflowed coordinates. These coordinates are constructed by using a novel and powerful MPNN that leverages the expressiveness of the attention mechanism. There exists yet another version of FermiNet, called PsiFormer, that uses the original self-attention mechanism introduced in~\cite{vaswani2017attention}, but has not been applied to extended systems. While PsiFormer applies the attention mechanism along the \emph{feature dimension} of the single-particle stream, we find substantially improved results by applying attention along the \emph{particle dimension} instead. This approach introduces additional correlations between the particles rather than between the features of the edge-states of our graphs. As PauliNet, we use a multiplicative factor to alter the orbitals even further, which allows us to omit the overall Jastrow factor that is present in PauliNet. By inputting orbital information to the multiplicative factor directly, we can keep the number of parameters independent of the number of particles. Notably, our Ansatz does not require multiple Slater determinants.

\begin{figure}[t]
  \includegraphics[width=\columnwidth]{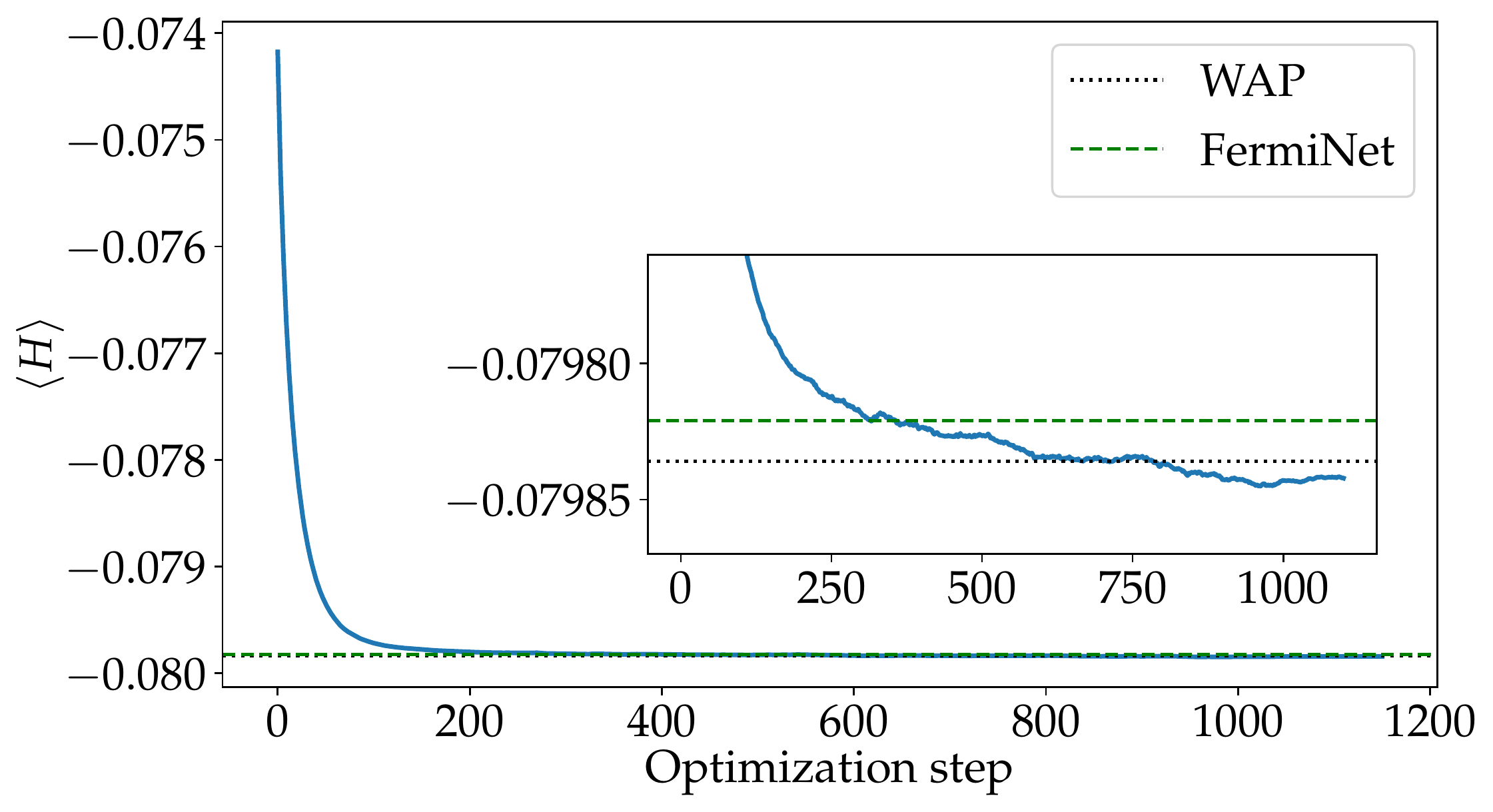}
  \caption{\label{fig::trainingcurve} Moving average of the optimization curve (energy per particle in Hartree averaged over 100 optimization iterations) for $N=14$ particles at $r_s=5$ obtained with a learning rate $\eta=0.05$ and a diagonal shift $\epsilon=10^{-4}$ for SR. We compare to results from WAP-net~\cite{wilson2022wave} and FermiNet~\cite{cassella2022discovering} to our MP-NQS. Note that the physically motivation orbitals used in our Ansatz makes us start close to the ground-state energy with our optimization.}
\end{figure}

As mentioned above, FermiNet has been applied to study the HEG in continuous space. The orbitals are constructed from scratch with plane-wave envelope functions. In contrast to our approach, FermiNet uses large networks (512 width) to process the positions of the simulated particles and considerably smaller networks (32 width) to act on the distance vectors between the particles. The usage of single-particle coordinates breaks the translation invariance of the system. In our MP-NQS, we only use the distance information between the particles and ignore the single-particle information to stay translation invariant. We only use small networks (32 width) to construct the messages $\mathbf{m}_{ij}^{(t+1)}$ (Eq.\ \ref{eq::messages}) that are used to update the nodes and edges of the particle graph. Overall, we use around 19000 parameters in our Ansatz which consists of a single Slater determinant. PauliNet usually uses 70000-100000 parameters and up to 36 Slater determinants, while FermiNet uses on the order of a million parameters with up to 32 Slater determinants. The fact that we have considerably fewer parameters allows us to use SR to optimize our variational Ansatz (see Appendix~\ref{Optimization}). Both FermiNet and PauliNet use $\mathcal{O}(10^5)$ optimization steps, but we can reduce this number to less than $\mathcal{O}(10^3)$ iterations. To illustrate this point, we show the optimization curve for $N=14$ particles at $r_s=5$ in Figure \ref{fig::trainingcurve}.

\section{Diffusion Monte Carlo\label{DMC}}
The diffusion Monte Carlo (DMC) method projects the variational states $|\Psi_V\rangle$ in imaginary time to filter out its ground-state component
\begin{equation}
    |\Psi_0\rangle \propto \lim_{\tau\rightarrow\infty} e^{- H\tau} |\Psi_V\rangle\, .
\end{equation}
The variational state is taken to be of the Slater-Jastrow form $\Psi_V = e^{J(\mathbf{X})} S_\uparrow S_\downarrow$, where $S_\uparrow$ and $S_\downarrow$ are the Slater determinants of single-electron states for the spin-up and spin-down species, respectively. The Jastrow factor captures correlation between electrons, and hence is a function of electron-electron separation coordinates $\mathbf{r}_{ij} = \{x_{ij}, y_{ij}, z_{ij}\}$
\begin{equation}
J(\mathbf{X}) = \sum_{\substack{i<j \\ \sigma, \tau \in \{\uparrow, \downarrow\}}}
J_{\sigma \tau}(\mathbf{r}_{ij})\,. \\
\end{equation}
The Pauli principle requires the functional form of $J_{\sigma \tau}$ must be even under exchange of particles $i$ and $j$. In addition, $J_{\sigma \tau}$ must satisfy the periodic-boundary conditions at the edge of the simulation box. To fulfill the above requirements, we utilize the parametrization introduced in Ref.~\cite{whitehead:2016}
\begin{equation}
J_{\sigma, \tau}(\mathbf{r}_{ij}) = \sum_{n=1}^{N_j} c_{n,\sigma\tau} [j(x_{ij})^2 + j(y_{ij})^2 + j(z_{ij})^2]^{n/2}\, ,
\end{equation}
where $j(x) = |x| [1-2(|x|/L)^3]$. By taking $c_{1,\uparrow\uparrow} = c_{1,\downarrow\downarrow} = \frac{1}{4}$ and $c_{1,\uparrow\downarrow} = c_{1,\downarrow\uparrow} = \frac{1}{2}$, the Kato cusp conditions at particle coalescence is also automatically fulfilled. The parameters $c_{n>1,\uparrow\uparrow}$ are determined by minimizing the energy of the systems by using the ``linear'' optimization scheme~\cite{Toulouse:2007}. We find that using more than $N_J = 6$ variational parameters does not improve the energy noticeably. 

The imaginary-time diffusion is broken into many small steps $\delta \tau$. The Trotter-Suzuki decomposition can be applied to the short-time propagator as $e^{-H\delta\tau} \simeq e^{-V \delta\tau / 2} e^{-T \delta\tau} e^{-V \delta\tau /2}$, where $T$ and $V$ are the kinetic and potential energy operators, respectively. At each imaginary-time step, we use the free propagator
\begin{align}
G_0(\mathbf{X}^\prime, \mathbf{X}) & \equiv  \langle \mathbf{X}^\prime | e^{-T \delta\tau} | \mathbf{X} \rangle  \nonumber \\
& = \Big[ \sqrt{\frac{m}{2 \pi \hbar^2  \delta \tau }}\Big]^{3N} \exp\Big[ -\frac{(\vec{R}^\prime -\vec{R})^2}{2\hbar^2\ \delta \tau/m} \Big]
\label{eq:free_prop}
\end{align}
to sample the new coordinates $\mathbf{X}^\prime$ of all particles. As routinely done in nuclear-physics applications~\cite{Pudliner:1997ck,Gandolfi:2020pbj}, to remove the linear terms coming from the exponential of Eq.~\eqref{eq:free_prop}, we use two mirror samples $\mathbf{X}^\prime = \mathbf{X} \pm \delta \mathbf{X}$, and we consider the corresponding importance-sampled weights
\begin{equation}
w_{\pm} = \frac{\Psi_V( \mathbf{X} \pm \delta \mathbf{X})} {\Psi_V( \mathbf{X})} e^{-[V(\mathbf{X} \pm \delta \mathbf{X}) + V(\mathbf{X})] \delta\tau / 2}\, .
\label{eq:weight_pm}
\end{equation}
Only one of the two samples is kept in the propagation according to a heat-bath sampling among the two normalized weights $w_{\pm}/(\sum_{\pm} w_{\pm})$ and the average weight $\sum_{\pm} w_{\pm}/2$ is assigned to the propagated configuration. 

The fermion-sign problem is controlled by employing the fixed-node approximation, which amounts to evaluating the weights of Eq.~\eqref{eq:weight_pm} with the replacement
\begin{equation}
\frac{\Psi_V( \mathbf{X} \pm \delta \mathbf{X})} {\Psi_V( \mathbf{X})} \to \Re{\frac{\Psi_V( \mathbf{X} \pm \delta \mathbf{X})} {\Psi_V( \mathbf{X})}}\, .
\end{equation}
Note that if the real part of the above ratio is negative, the weight of the configuration is set to zero only after computing the average weight.

\section{Results \label{app::results}}
In Tables \ref{tab::unpolarized14}, \ref{tab::unpolarized54}, \ref{tab::polarized19} and \ref{tab::polarized27} we show the ground-state energies of the unpolarized and fully polarized HEG for different system sizes $N=\{14,19,27,54\}$ and densities $r_s=\{1,2,5,10,20,30,50,70,90,100\}$.
\begin{table*}[ht]
    \centering
    \begin{tabular}{c|c c c c c}
    \hline \hline
        N & $r_s$ & MP-NQS & WAP~\cite{wilson2022wave} & FermiNet~\cite{cassella2022discovering} & FCI$^\ast$/DCD$^{\ast\ast}$~\cite{alavi2021benchmark} \\ \hline
        14 & 1 & $0.568967(6)$ & $0.568965(1)$ & $0.568904(1)$ & $0.56861(1)^\ast$  \\
        ~ & 2 & $-0.008391(1)$ & $-0.0083310(3)$& $-0.008427(1)$ & $-0.00868(2)^\ast$ \\
        ~ & 5 & $-0.0798544(4)$ & $-0.0798360(1)$ & $-0.079821(1)$ & $-0.08002(2)^\ast$ \\
        ~ & 10 & $-0.0552126(6)$ & $-0.05520380(3)$ & N/A & $-0.05509^{\ast\ast}$ \\ 
        ~ & 20 & $-0.0324553(2)$ & $-0.0324434(1)$ & N/A & $-0.03201^{\ast\ast}$\\ 
         ~ & 50 & $-0.01462631(6)$ & $-0.01462211(4)$ & N/A & $-0.01384^{\ast\ast}$ \\ 
         & 100 & $-0.00773018(3)$ & $-0.007729980(2)$ & N/A & N/A \\  \hline \hline
    \end{tabular}
    \caption{\label{tab::unpolarized14}Total energy per particle in Hartree for unpolarized system of $N=14$ particles. WAPNet and FermiNet are alternative NQS architectures optimized via VMC. We include FCI and DCD results as benchmarks from quantum chemistry.}
\end{table*}
\begin{table*}[ht]
    \centering
    \begin{tabular}{c|c c c c c c c }
    \hline \hline
        N & $r_s$ & MP-NQS & MP-NQS (Gauss) & LiNet~\cite{li2022ab} & FN-DMC  & BF-DMC~\cite{PhysRevE.74.066701,PhysRevB.105.245135} & FCI$^\ast$/DCD$^{\ast\ast}$~\cite{alavi2021benchmark} \\ \hline
        54 & 1 & $0.52973(1)$ & N/A & $0.530019(1)$ & $0.53094(2)$ & $0.52989(4)$ & $0.52973(3)^\ast$ \\
        ~ & 2 & $-0.014046(8)$ &  $-0.01390(1)$ & $-0.013840(1)$ & $-0.01326(2)$ & $-0.013966(2)$ & $-0.01379^{\ast\ast}$ \\ 
        ~ & 5 & $-0.079090(2)$ & $-0.079064(4)$ & $-0.0788354(2)$ & $-0.07867(1)$ & $-0.079036(3)$ & $-0.07837^{\ast\ast}$ \\
        ~ & 10 & $-0.054448(1)$ & $-0.054445(1)$ & $-0.0542785(1)$ & $-0.054269(8)$ & $-0.054443(2)$& $-0.05322^{\ast\ast}$ \\ 
        ~ & 20 & $-0.0320524(5)$ & $-0.0320480(6)$ & $-0.0316886(1)$ & $-0.031976(8)$ & $-0.032047(2)$ & $-0.03113^{\ast\ast}$ \\ 
        ~ & 50 & $-0.0145015(1)$ & $-0.0144999(1)$ & N/A & $-0.01387(2)$ & $-0.0144877(1)$ &$-0.01281^{\ast\ast}$ \\ 
        ~ & 100 & $-0.0076793(1)$ & $-0.00769203(5)$ & N/A & $-0.007674(3)$ & N/A & N/A \\  \hline \hline
    \end{tabular}
    \caption{\label{tab::unpolarized54}Total energy per particle in Hartree for the unpolarized system of $N=54$ particles. FN-DMC results are obtained using the method in Appendix~\ref{DMC}. We include FCI and DCD results as benchmarks from quantum chemistry.}
\end{table*}
\begin{table}[t]
    \centering
    \begin{tabular}{c|c  c c c}
    \hline \hline
        N & $r_s$ & MP-NQS & MP-NQS (Gauss) & FN-DMC  \\ \hline
        128 & 110 & $-0.0070203(1)(1)$ & $-0.0071121(2)$ & $-0.007009518(1)$  \\ 
        ~ & 200 & $-0.0040023(2)$ & $-0.0040615(1)$ & $-0.0039923(7)$ \\
         \hline \hline
    \end{tabular}
    \caption{\label{tab::unpolarized128}Total energy per particle in Hartree for the unpolarized system of $N=128$ particles. FN-DMC results are obtained using the method in Appendix~\ref{DMC}.}
\end{table}
\begin{table}[t]
    \centering
    \begin{tabular}{c|c c c}
    \hline \hline
        N & $r_s$ & MP-NQS & WAP~\cite{wilson2022wave}   \\ \hline
        19 & 1 & $1.046262(2)$ & $1.046241(3)$  \\ 
        ~ & 2 &  $0.096307(1)$ & $0.096303(1)$  \\ 
        ~ & 5 & $-0.0672489(3)$ & $-0.06725105(3)$ \\ 
        ~ & 10 & $-0.0528624(2)$ & $-0.05286035(1)$ \\ 
        ~ & 20 & $-0.0320101(1)$ & $-0.0320082(1)$ \\ 
        ~ & 50 & $-0.01456685(3)$ & $-0.01456571(2)$ \\ 
        ~ & 100 & $-0.00771832(2)$ & $-0.00771362(2)$ \\ 
         \hline \hline
    \end{tabular}
    \caption{\label{tab::polarized19}Total energy per particle in Hartree for the polarized system of $N=19$ particles. We compare to results obtained using WAPNet.}
\end{table}
\newpage
\begin{table}[t]
    \centering
    \begin{tabular}{c|c c c c}
    \hline \hline
        N & $r_s$ & MP-NQS & FN-DMC  \\ \hline
        27 & 1 & $1.051771(4)$ & $1.05200(3)$  \\ 
        ~ & 2 & $0.098460(2)$  & $0.09866(2)$ \\ 
        ~ & 5 & $-0.0665523(6)$ & $-0.066429(4)$  \\ 
        ~ & 10 & $-0.0526015(2)$ & $-0.052540(6)$  \\ 
        ~ & 30 & $-0.0228184(1)$ & $-0.0228080(8)$ \\ 
        ~ & 50 & $-0.01457976(5)$ & $-0.0145752(6)$  \\ 
        ~ & 70 & $-0.01074986(3)$ & $-0.010749(2)$ \\ 
        ~ & 90 & $-0.00853156(2)$ & $-0.0085314(5)$\\ 
         \hline \hline
    \end{tabular}
    \caption{\label{tab::polarized27}Total energy per particle in Hartree for the polarized system of $N=27$ particles. FN-DMC results are obtained using the method in Appendix~\ref{DMC}.}
\end{table}

\end{document}